\documentclass[final,5p,times,twocolumn]{elsarticle}

\usepackage{lineno}
\usepackage[hidelinks]{hyperref}
\usepackage{multirow}
\usepackage{graphicx}
\usepackage{amsmath}   
\usepackage{float}
\usepackage[section]{placeins}

\begin{document}

\begin{frontmatter}

\title{PSEC6: an 8-Channel 40 GSa/s Waveform Sampling ASIC in TSMC 65nm with 10.24 GHz PLL}

\author[1]{Ahan Datta\corref{cor1}}
\ead{ahand@uchicago.edu}
\author[1]{Andrew Arzac}
\author[2]{Davide Braga}
\author[1]{Gordon Chen}
\author[2]{Troy England}
\author[2]{Farah Fahim}
\author[1]{Henry J. Frisch}
\author[1]{Nathan Gehl}
\author[1]{Mary Heintz}
\author[1]{Sumin Kim}
\author[1]{Ava Lalich}
\author[7]{Jinseo Park}
\author[2]{Nathaniel J. Pastika}
\author[6]{Hector D. Rico-Aniles}
\author[2]{Paul M. Rubinov}
\author[2]{Xiaoran Wang}
\author[1]{Y.M. Richmond Yeung}
 \cortext[cor1]{Corresponding author}

\affiliation[1]{organization={Enrico Fermi Institute, the University of Chicago}, 
                 addressline={933 East 56th Street},
                 postcode={60637}, 
                 city={Chicago, IL}, 
                 country={USA}}
\affiliation[2]{organization={Fermi National Accelerator Laboratory},
                 postcode={60510}, 
                 city={Batavia, IL}, 
                 country={USA}}

\affiliation[6]{organization={North Central College}, 
                 addressline={30 N. Brainard Street},
                 postcode={60540}, 
                 city={Naperville, IL}, 
                 country={USA}}

\affiliation[7]{organization={Stanford University}, 
                 addressline={450 Jane Stanford Way},
                 postcode={94305}, 
                 city={Stanford, CA}, 
                 country={USA}}

\begin{abstract}
Picosecond level timing resolution is a prerequisite capability for improved coincidence matching, time-of-flight measurements, and secondary vertex reconstruction.
Here, we present the specification, design, and simulation results for a new Application Specific Integrated Circuit (ASIC), called PSEC6, in the TSMC 65nm process.
It features 8 channels, a maximum sampling rate of 40 GSa/s, a buffer length of 204.8 nanoseconds, and a 10.24 GHz Phase Locked Loop (PLL), which is the first of its kind in the 65nm CMOS process.
The event readout rate is 32 kHz, with the digitization done by an off-chip Analog-to-Digital Converter (ADC).

Simulations predict a 4.0 GHz analog input bandwidth and 20 mW per channel during sampling; the 10.24 GHz PLL has a predicted jitter of 550 fs RMS at 15.7 mW.
The paper describes the sampling architecture, chip signal paths, PLL design, and presents simulation results.
\end{abstract}


\end{frontmatter}


\section{Introduction}

Single picosecond timing resolution is key to implementing high accuracy time-of-flight measurements and secondary/tertiary vertex reconstruction at colliders.
For many applications, the readout electronics must be multi-hit capable, have a long buffer, a high sampling rate, a high analog bandwidth, and a low power consumption \cite{ritt2011analog}.
These criteria define the specifications of PSEC6, as given in Table \ref{tab: PSEC6 specifications}.

PSEC6 is an extension of the PSEC5 ASIC which was taped out in August 2024 \cite{park2025psec5}.
For PSEC6, we add a 10.24 GHz PLL to reduce timing drift across sampling runs.
We additionally re-designed the clock divider chain and digital blocks to eliminate failure modes in the slow process corner.

PSEC6 is expected to achieve single picosecond timing resolution with a 40 GSa/s sampling rate and a 4 GHz analog bandwidth.
Simulations predict average power consumption of each of the eight channels to be 20 mW and of the PLL to be 15.7 mW.

PSEC6 relies on an external ADC to convert the sampled capacitor voltages to digital data.
This reduces chip complexity, power cost, and die size while allowing for flexibility in the choice of ADC.
Using a 40 MHz ADC, we estimate a dead time per event of 32 $\mu$s.

\begin{figure*}[t]
    \centering
    \includegraphics[width=\linewidth]{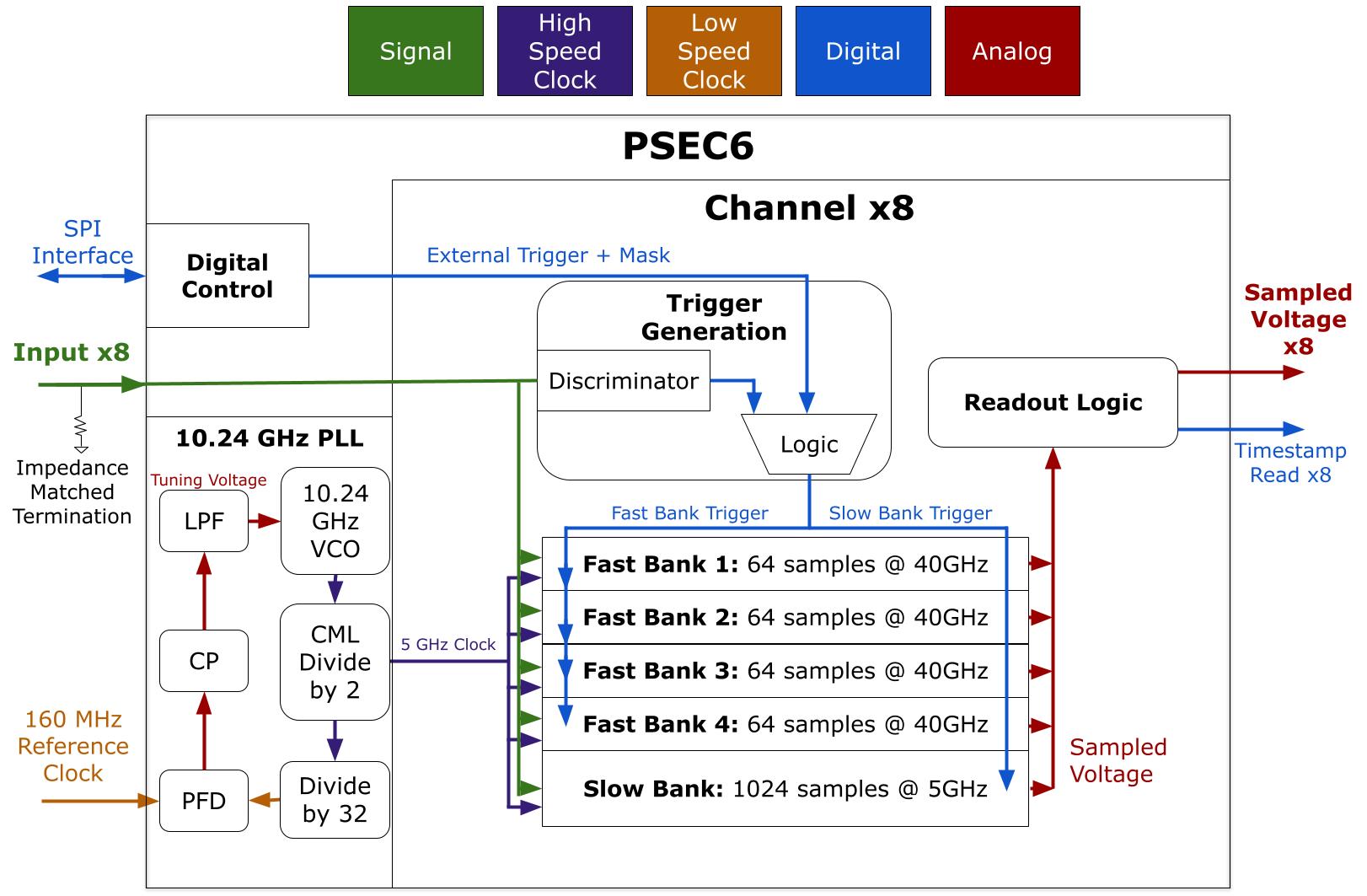}
    \caption{\textbf{Block diagram for PSEC6}:
    The layout and different sampling frequencies of the fast and slow buffer switched capacitor array (SCA) architectures are shown. 
    The fast banks are triggered by the per-channel trigger generation logic.
    The sampling clock is generated by the 10.24 GHz PLL.
    The analog readout is controlled by the digital readout logic and fed into external ADCs for digitization.
    The initialisms in the PLL are defined in Section \ref{sec: PLL}.}
    \label{fig: PSEC6 block diagram}
\end{figure*}

The block diagram of the chip is given in Figure \ref{fig: PSEC6 block diagram}.
Each detector input is buffered through to the sampling switches.
The sampling switches are triggered by the sampling clock at 10 GHz, driven from the PLL. 
The sampled voltages are read out using a further pair of source followers to external ADCs. 
The various settings on the chip are controlled by the digital block \cite{datta2024cpad}.

\begin{table}[h]
\centering
    \begin{tabular}{|l|l|}
    \hline
    Process & 65nm TSMC \\ \hline
    Signal to Noise & 1000 \\ \hline
    \textbf{Sampling Rate} & \textbf{40GSa/s (5GSa/s)} \\ \hline
    Fast Buffer Length & 4$\times$1.6 ns \\ \hline
    Slow Buffer Length & 204.8ns \\ \hline
    Analog Bandwidth & 4.9 GHz \\ \hline
    Power Consumption (Sampling) & 20 mW (per channel) \\ \hline
    Channels/Chip & 8 \\ \hline
    Area & 2.7 mm$^2$ \\ \hline
    Readout Rate & 32 kHz \\ \hline
    \end{tabular}
\caption{Specifications for PSEC6 ASIC}
\label{tab: PSEC6 specifications}
\end{table}

\section{Switched Capacitor Array}

Switched Capacitor Arrays (SCA) are sequentially-triggered capacitor-based sample-and-hold circuits which sample the voltage input at a frequency determined by the input clock frequency.
Given that each sample captures a certain time segment and that the number of sampling capacitors per channel is fixed due to size considerations, there is an inverse relationship between sampling rate and buffer length.

\subsection{Fast and Slow Bank Architecture}

To achieve both a high sampling rate and a long sampling window, we employ two different types of SCAs: fast banks and slow banks.
In each channel, there are four fast banks and one slow bank.
The slow bank samples at 5 GSa/s over a buffer length of 204.8 nanoseconds.
The four independent fast banks sample at 40 GSa/s with a shorter buffer length of 1.6 nanoseconds each.
The fast banks are triggered by a discriminator whose parameters are controlled by the user \cite{park2025psec5}.

To achieve a 40 GSa/s sampling rate from a 10 GHz clock, the fast buffers employ a four-phase interleaved sampling scheme.
The sampling clock incurs four $90^\circ$ phase offsets within each fast buffer.
Thus, the fast buffer can sample at 40 GSa/s while being driven by a single 10 GHz clock.
The 10 GHz clock is divided once to provide the 5 GHz sampling clock for the slow buffer.

The user has the choice to either run the fast buffers sequentially or to allow each fast buffer to run independently. In the former case, there is a 6.4 nanosecond sampling window for a single event.
In the latter case, the chip is multi-hit capable for up to four distinct events, with a sampling window of 1.6 nanoseconds per event.
PSEC6 can also be configured to sample two distinct events with a sampling window 3.2 nanoseconds per event.

The slow buffer fulfills two purposes.
For one, it samples the baseline of the detector, allowing for baseline calibration and noise characterization.
Additionally, it serves as a long time series in which to temporally locate the events captured in the fast bank.
This is especially important for applications that rely on time differences, such as time-of-flight.

\subsection{Switch Design}

\begin{figure}[tbp]
    \centering
    \includegraphics[width=\linewidth]{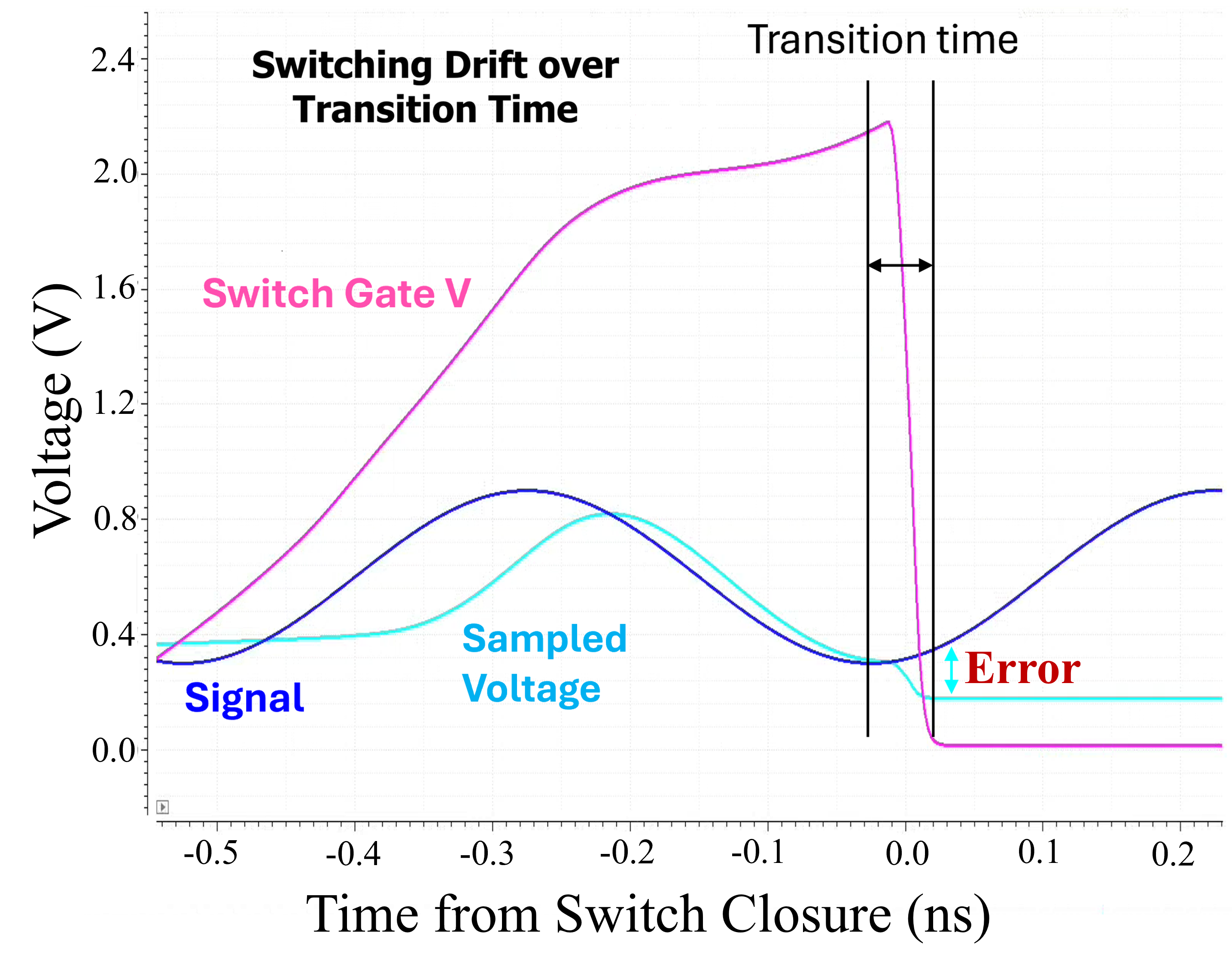}
    \caption{\textbf{Simulation of the Switching Drift}: SPICE Simulation of the switching drift induced by a finite gate voltage (light pink) transition time on the order of $0.1$ nanoseconds.
    The sampled voltage (light blue) drifts away from the true signal (dark blue) during the transition window, which adds a systematic error into the sample.}
    \label{fig: switching drift}
\end{figure}

The choice of sampling switch is important for three characteristics: analog bandwidth, physical gate length, and transition time.
High analog bandwidth is important to resolve the rising edge and separation between pulses with short rise times.
Physical gate length is a tradeoff between larger process variation at shorter lengths and larger area requirements at longer lengths.
A low transition time is important to reduce charge leakage after sampling, which induces irreducible systematic error (see Figure \ref{fig: switching drift}).

\begin{figure}[tbp]
    \centering
    \includegraphics[width=\linewidth]{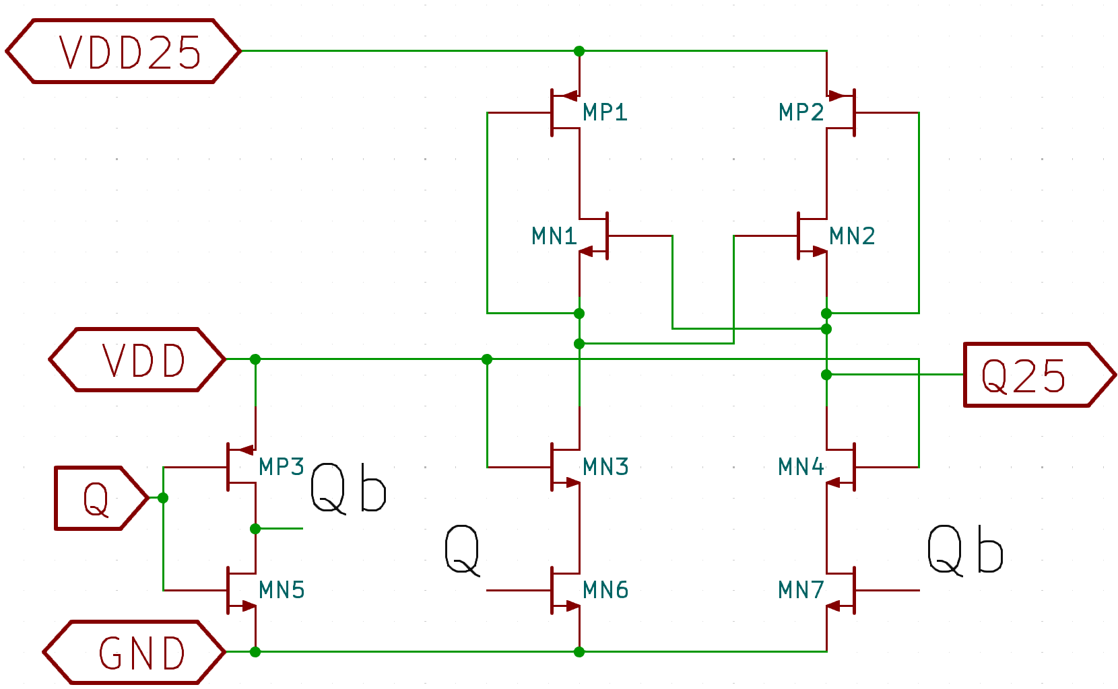}
    \caption{\textbf{Shifting the Gate Voltage}: Schematic of voltage level shifter required to convert the 1.2V CMOS sampling clock to the 2.5V gate voltage required for the NMOS I/O voltage sampling switch. }
    \label{fig: level shifter}
\end{figure}

As a compromise between these three characteristics, we chose to use an I/O voltage (2.5 V) NMOS transistor with 280 nanometer gate length.
The I/O voltage NMOS switch has a 6 GHz bandwidth and a hold time exceeding 20 $\mu$s at the cost of a longer gate length.
Our clock voltage, however, is 1.2 V, meaning that we need a 1.2 V to 2.5 V voltage level shifter.
The schematic used is in Figure \ref{fig: level shifter}.

\subsection{Clock Distribution}

A key input to the SCA is the sampling clock which triggers the sequential sample-and-hold circuits. 
We need the sampling clock to be low jitter with a near 50\% duty cycle.
Furthermore, the fast banks must see a sampling clock frequency of 10 GHz.
Unfortunately, parasitic elements of the clock distribution network make the distribution of a 10 GHz clock across the chip technically difficult and power intensive.
Thus, we opt to distribute a 5 GHz clock across the chip instead of the 10 GHz clock.

\begin{figure*}[t]
    \centering
    \includegraphics[width=\linewidth]{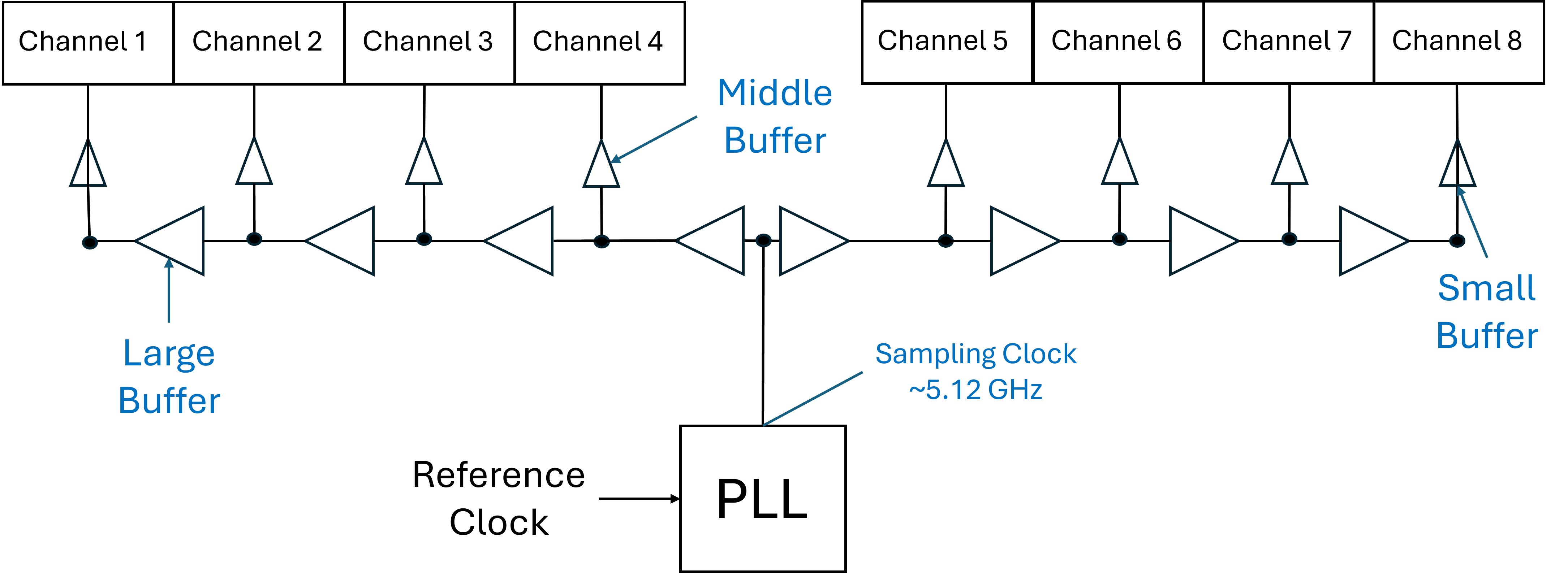}
    \caption{\textbf{Clock Distribution Buffer Chain}: Block diagram description of the clock distribution buffer chain. Each buffer is an inverter chain of four.
    The transistor widths of the buffers are in a ratio of 4:2:1.}
    \label{fig: buffer chain}
\end{figure*}

\begin{figure}[tbp]
    \centering
    \includegraphics[width=\linewidth]{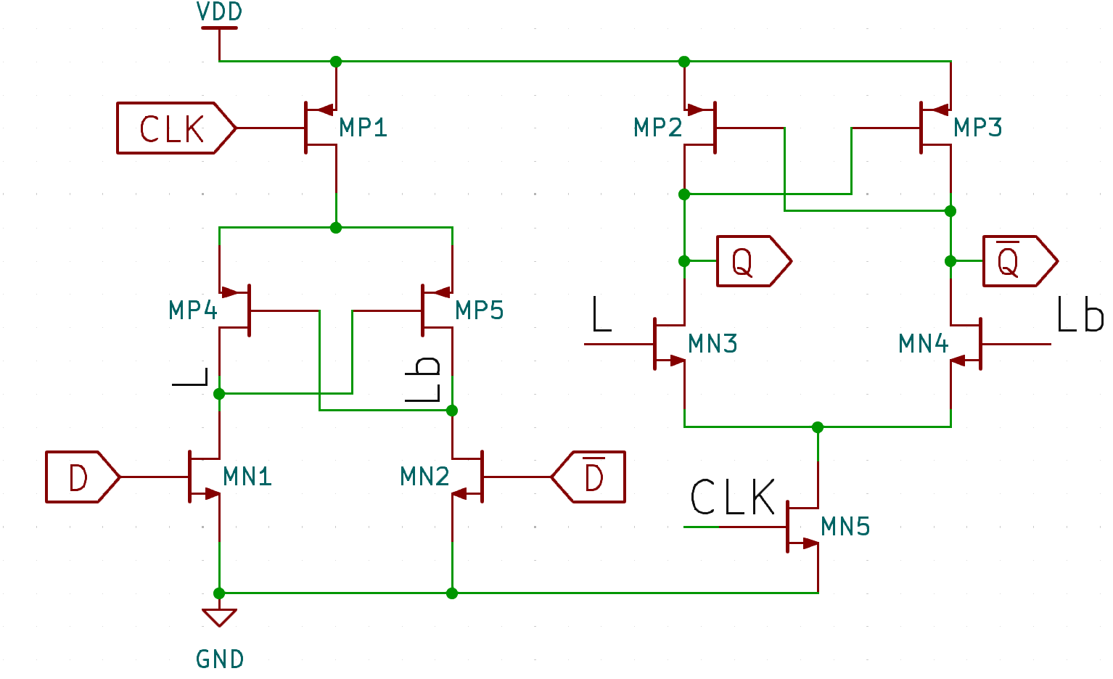}
    \caption{\textbf{Doubling the Clock Frequency}: Schematic of the dual-edge-triggered flip-flop (DET FF) used to reduce the distributed clock frequency to 5 GHz while maintaining the sampling frequency of 40 GSa/s.}
    \label{fig: DET FF}
\end{figure}

The clock distribution is handled by the buffer chain as shown in Figure \ref{fig: buffer chain}.
At each channel, 5 GHz distributed clock is converted to the 10 GHz sampling clock using an existing design of a Dual-Edge-Triggered Flip Flop (DET FF), as shown in Figure \ref{fig: DET FF} \cite{diffFF}.
The typical problems of dual-edge triggering, high power consumption and instability, are avoided by careful simulation of the duty cycle and jitter tolerances across the process variations likely to be seen by the chip.

\section{Signal Paths}

The critical signal paths for the chip are the input path and the analog readout path.
The input path is defined to be the path between the package pin input for the detector voltage and the sampling switch.
The readout path is defined to be the path from the sampling capacitor to the package pin output of the sampled voltage.

\subsection{Input Signal Path}

\begin{figure}[tbp]
    \centering
    \includegraphics[width=\linewidth]{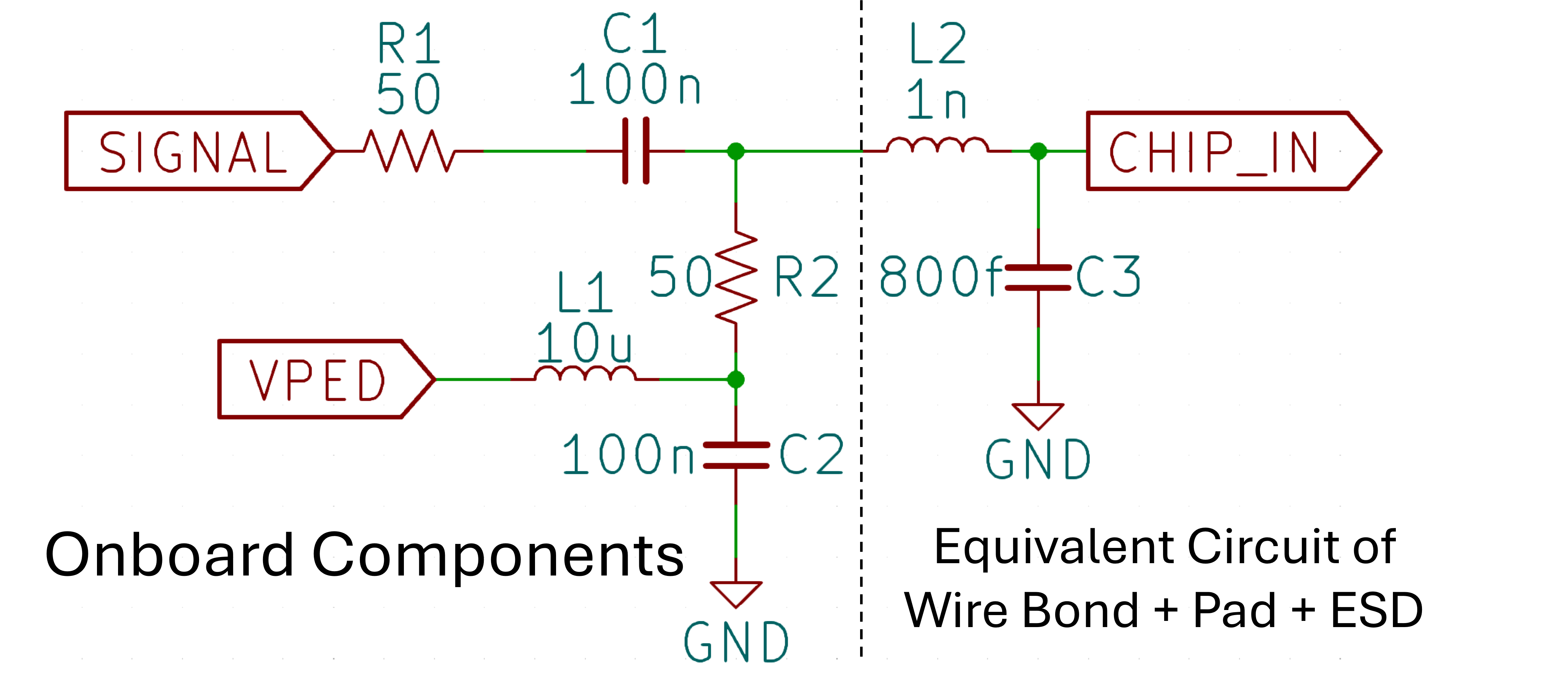}
    \caption{\textbf{Chip Entry Path}: Schematic of expected on-board components and parasitic components, arising from the ESD protection and pad devices on the chip, used for SPICE verification of the bandwidth.
    The pedestal voltage, $V_\text{ped}$, is expected to be supplied externally to bias the detector signal into the optimal range for the input buffer and discriminator.}
    \label{fig: chip in}
\end{figure}

The input path is characterized by the components shown in Figure \ref{fig: chip in}.
The onboard components will be on the PCB (Printed Circuit Board) which carries the chip, provides the biases, and AC-couples the detector output to the chip.
The components shown on the right are parasitics due to the wirebond, input pad, and ESD protection devices.
A SPICE simulation of the schematic returned an analog bandwidth of 4.0 GHz.

\begin{figure}[tbp]
    \centering
    \includegraphics[width=0.86\linewidth]{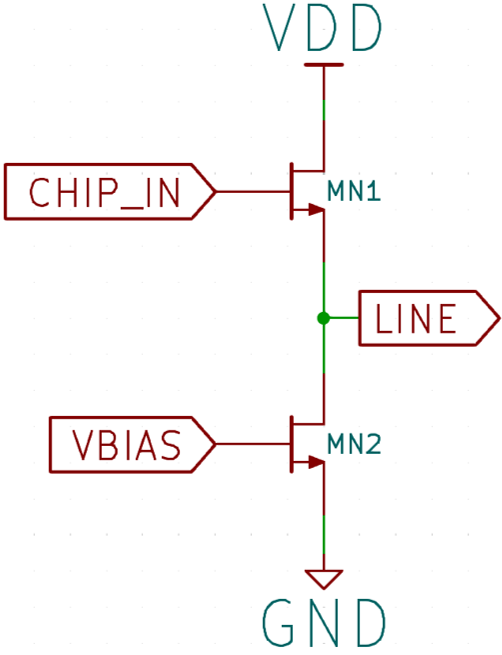}
    \caption{\textbf{Input Buffer for the Signal}: Schematic of the input source follower for the detector signal. 
    The bottom NMOS acts as a current source, using a global voltage bias, and the top NMOS takes the signal input.
    The bandwidth of this source follower was simulated to be 4.9 GHz. }
    \label{fig: input source follower}
\end{figure}

To reduce the loading on the detector output, we employ a source follower as shown in Figure \ref{fig: input source follower}.
The simulated analog bandwidth of the source follower was found to be 4.9 GHz.
Thus, the total analog bandwidth of the input signal path is 4.0 GHz.

\subsection{Readout}

The readout of the chip is controlled by a 40 MHz readout clock.
At each rising edge of the readout clock, the voltage from the next sampling capacitor in each channel is sent to the corresponding readout pin of the chip.
To reduce the charge leakage from the small sampling capacitors (35 fF), the output is buffered through two source followers, similar to the schematic in Figure \ref{fig: input source follower}, directly to the pad.
The output to the pad is analog and is digitized by an external ADC located on the PCB.

\section{Phase Locked Loop} \label{sec: PLL}

A Phase Locked Loop (PLL) is a feedback loop in phase space to lock the phase of an oscillator to that of an external reference \cite{wang2024cryopll, braga2024cryopll}.
Figure \ref{fig: PSEC6 block diagram} shows a block diagram of the PLL.

The PLL architecture is as follows.
The 10.24 GHz output of the oscillator is first divided down by 64, creating a 160 MHz PLL clock.
The PLL clock is compared against a 160 MHz reference clock, supplied by a crystal oscillator, to find the phase error by the Phase Frequency Detector. 
This phase error is then converted to a correcting voltage using a Charge Pump and Loop Filter, closing the loop \cite{datta2026aps}.

\subsection{Voltage Controlled Oscillator}

\begin{figure}[tbp]
    \centering
    \includegraphics[width=\linewidth]{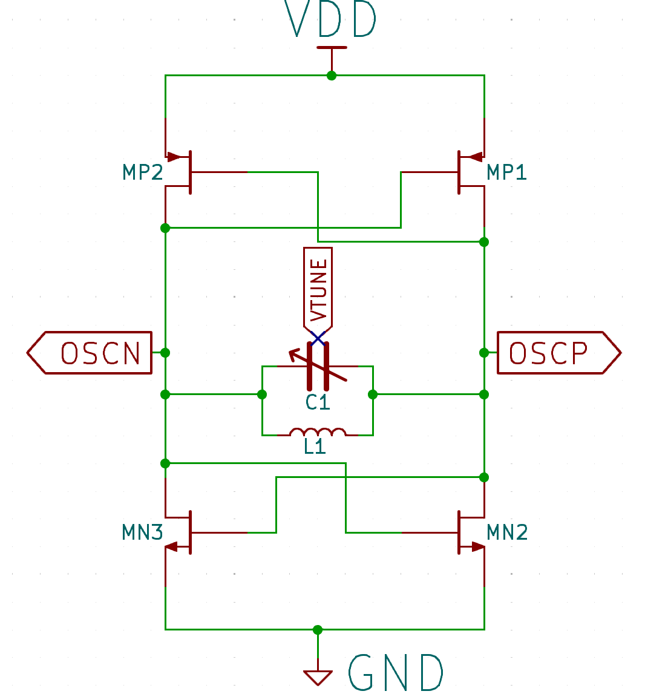}
    \caption{\textbf{Voltage Controlled $LC$-Oscillator}: Schematic of the LC-oscillator based Voltage Controlled Oscillator (VCO) used for the PLL. 
    The capacitor and inductor values are tuned to have a central frequency value of 10.24 GHz.
    The capacitance value is voltage-controlled using both an analog varactor and a digital switched capacitor bank.}
    \label{fig: vco schematic}
\end{figure}

The Voltage Controlled Oscillator (VCO) for this PLL, shown in Figure \ref{fig: vco schematic}, is a 10.24 GHz center frequency $LC$-VCO.
The VCO has a digital binary switched capacitor bank to adjust the frequency range.
The VCO also has an analog control voltage, $V_\text{tune}$, with a gain parameter $K_\text{VCO} \approx 220 \text{ MHz/V}$, and has a maximum tuning range of 9.75 GHz to 11.04 GHz.
The PLL is implemented using the analog feedback, $V_\text{tune}$. 

\subsection{Divider Chain}

\begin{figure*}[t]
    \centering
    \includegraphics[width=\linewidth]{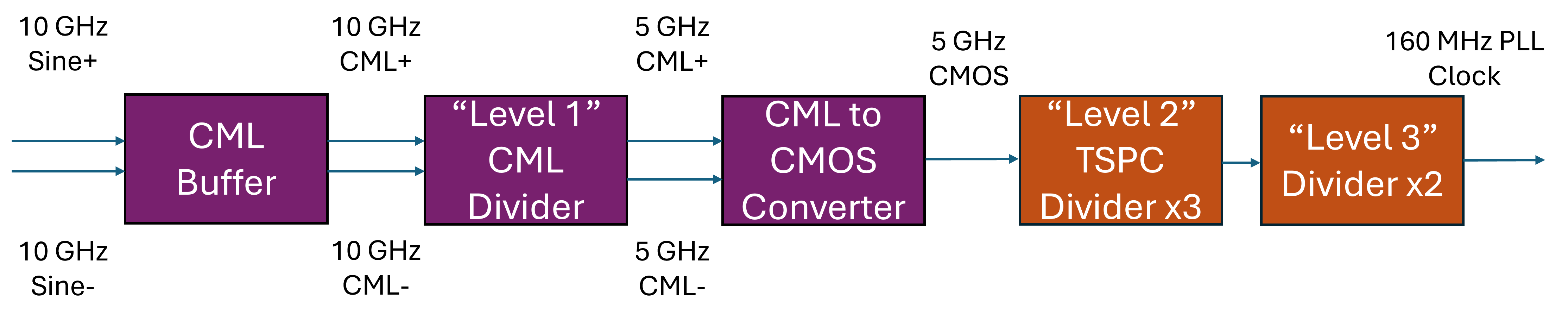}
    \caption{\textbf{PLL Clock Divider Chain}: Block diagram description of the divider chain used for the PLL feedback, sampling clock, and output clock paths.
    The first stage, shown in purple, uses Current-Mode Logic (CML), trading high power consumption for 10.24 GHz division.
    The later stages, which operate at lower frequencies, are shown in orange.
    The second level dividers uses True Single Phase Clock (TSPC) logic, balancing power consumption and speed.
    The third level dividers are standard single flip-flop dividers for minimal power consumption.}
    \label{fig: divider chain}
\end{figure*}

The divider chain converts the 10.24 GHz output of the VCO into the 5.12 GHz sampling clock and the 160 MHz PLL clock, as shown in Figure \ref{fig: divider chain}. 
The first level of divider is implemented in Current-Mode Logic (CML), which uses more power and can operate at higher frequencies. 
The output of the first level divider is used for the 5.12 GHz sampling clock.
The CML stage designs were derived from References \cite{heydari2003cml, mohanavelu2004divider}.
The second level of dividers is built in True-Single-Phase-Clock logic (TSPC logic), striking a balance between power consumption and operating frequency \cite{razavi2016tspc}.
The third and final level of divider is constructed using a single flip-flop divider to lower the power consumption.
The output of the third level dividers is used for the 160 MHz PLL clock.

\subsection{Phase Frequency Detector}

\begin{figure}[tb]
    \centering
    \includegraphics[width=\linewidth]{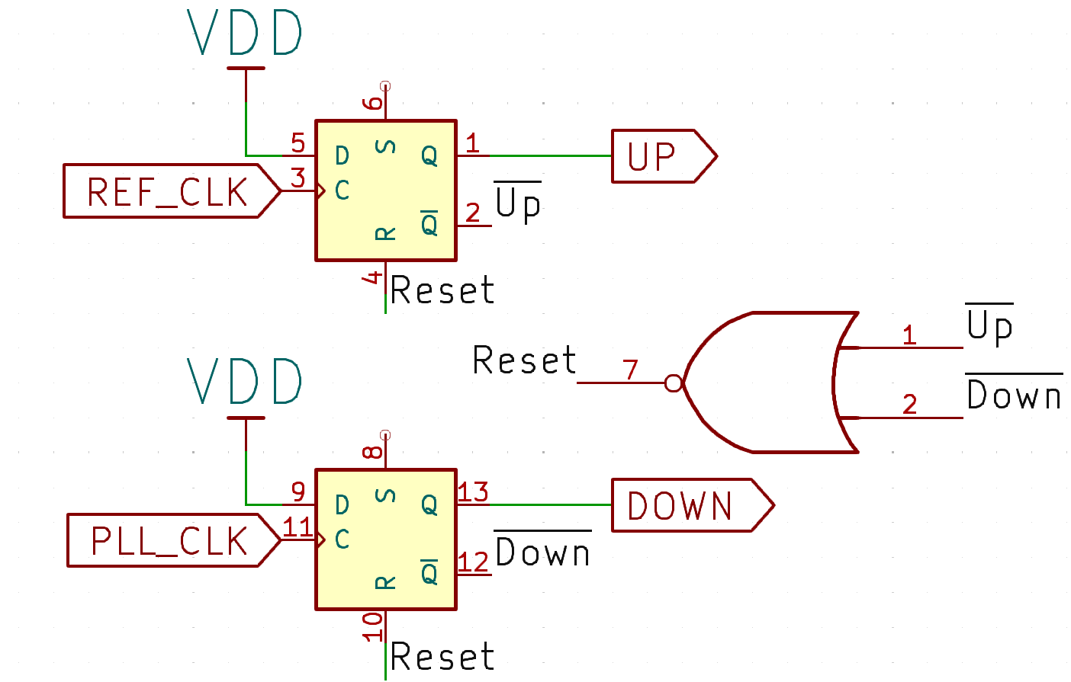}
    \caption{\textbf{Phase Frequency Detector}: Schematic of the two-flip-flop Phase Frequency Detector (PFD) used in the PLL.
    The feedback path is designed around a NOR gate to minimize the reset path time, which improves stability and performance of the PLL.}
    \label{fig: pfd}
\end{figure}

The Phase Frequency Detector (PFD) used in this PLL is shown in Figure \ref{fig: pfd}.
Lower reset times improve the stability and locking time of the PLL by decreasing the total feedback path delay of the PLL \cite{mansuri2002pfd, sharma2021pfdreview}.
Thus, we use a \texttt{NOR} gate in the reset path to minimize the reset time of the PFD.
This has the added benefit of symmetrically loading the two D-flip-flops used in the PFD, which reduces the magnitude of reference spurs caused by PLL asymmetry.  

\subsection{Charge Pump}

\begin{figure}[tb]
    \centering
    \includegraphics[width=\linewidth]{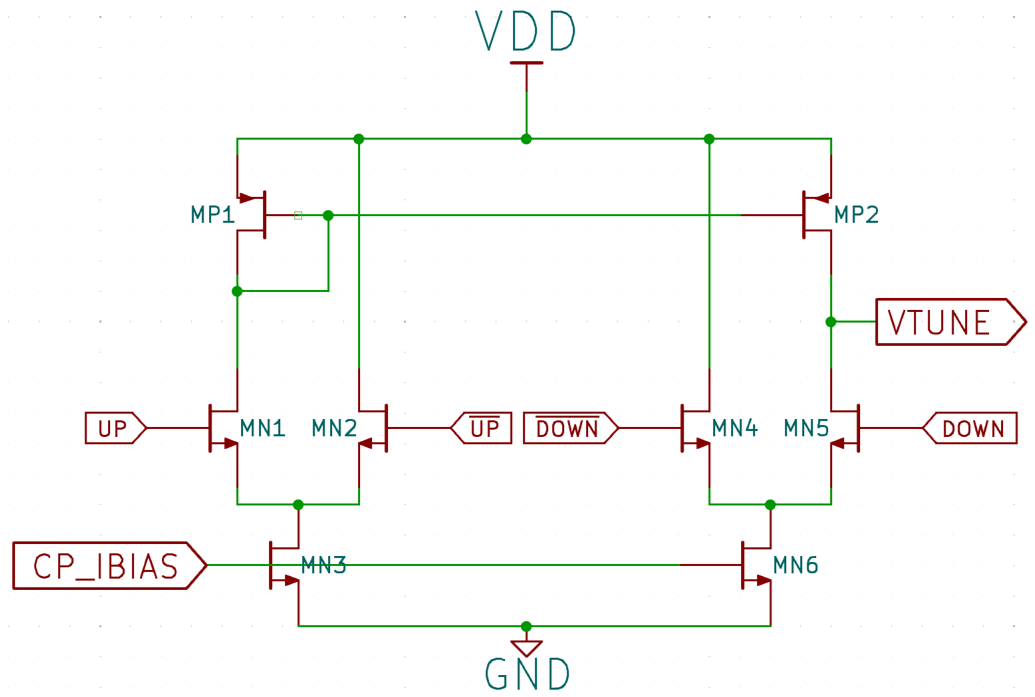}
    \caption{\textbf{Charge Pump}: Schematic of the symmetric Charge Pump (CP) used in the PLL.
    The use of primarily NMOS transistors for both the up and down path reduces the asymmetry in the feedback paths, as it avoids the inherent characteristic difference between PMOS and NMOS transistors.
    The current mirror required for the up path induces a minor asymmetry in the design, hurting performance.}
    \label{fig: charge pump}
\end{figure}

The Charge Pump (CP) design used in the PLL is given in Figure \ref{fig: charge pump}.
This charge pump is a current-steering design using NMOS steering transistors for symmetry.
The current sink bias is controllable by the user to determine the right balance of phase margin, power consumption, and settling time.
The current mirror in the up path introduces some unavoidable asymmetry.
This CP was modeled from an existing design \cite{rhee1999chargepump}.

\subsection{Loop Filter}

\begin{figure}[tb]
    \centering
    \includegraphics[width=\linewidth]{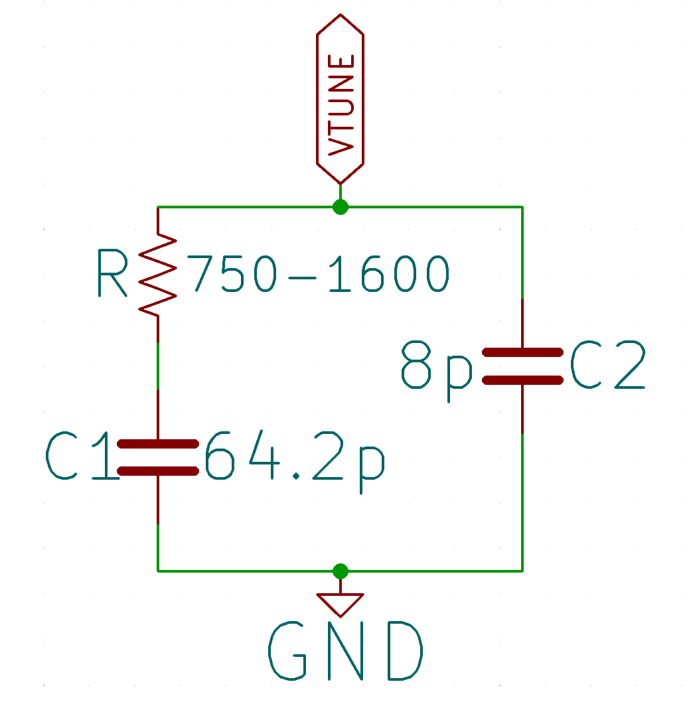}
    \caption{\textbf{Loop Filter}: Schematic of the Loop Filter (LPF) used in the PLL.
    The exact capacitor and resistor values, shown in units of Farads and Ohms respectively, were determined to reduce the loop jitter while maintaining an adequate phase margin.
    $C_2$ is kept an order of magnitude smaller than $C_1$ such that we are able to approximate the PLL as type-II with a simpler pole structure.
    The resistor is variable using a switched resistor array controlled digitally.}
    \label{fig: loop filter}
\end{figure}

The Loop Filter (LPF) used in this PLL is shown in Figure \ref{fig: loop filter}.
The LPF smooths the output of the charge pump and determines the phase margin and poles of the PLL.
The LPF component sizing controls the phase margin, which is critical to the PLL stability. 
Process variation can shift the value of $R$ by up to 20\%.
Thus, we implemented a tunable resistor between $750 \ \Omega$ and $1.6 \ \text{k}\Omega$ to ensure that the PLL is always stable.

\subsection{PLL Simulation Results}

\begin{figure*}[t]
    \centering
    \includegraphics[width=\linewidth]{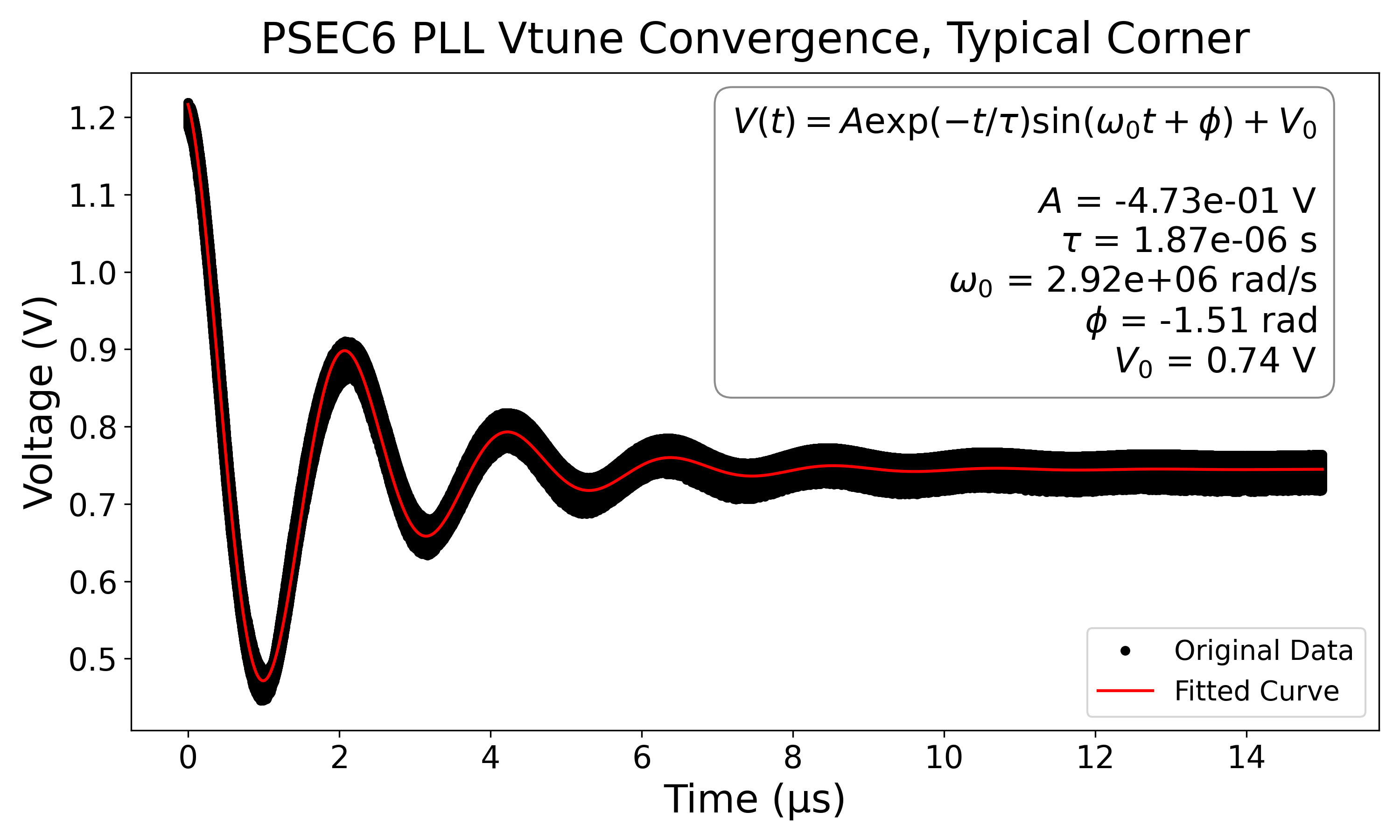}
    \caption{\textbf{PLL $V_\text{tune}$ Convergence}: Plot showing the convergence of $V_\text{tune}$, the tuning voltage of the VCO, in time as a function of the PLL.
    This simulation is done in the typical corner.
    Using an exponentially decaying sinusoidal fit, we extract a settling voltage, $V_0$, near mid-rail (0.7446 V) and a natural frequency, $\omega_0$, in agreement with the calculated value.}
    \label{fig: vtune convergence}
\end{figure*}

The results of simulating the PLL in the typical process corner are shown in Figure \ref{fig: vtune convergence}.
The convergence voltage, $V_0$, being near mid-rail implies that the PLL frequency and VCO center frequency match.
The convergence timescale, $\omega_0$, is on the order of microseconds, which meets the desired specifications. 
The detailed PLL specifications across corners is given in Table \ref{tab: PLL specifications}.

\begin{table}[h]
\centering
\begin{tabular}{|l|c|}
\hline
\textbf{Figure of Merit} & \textbf{Value} \\ \hline
Frequency & 10.24 GHz \\ \hline
Average Total Power & $P_{\text{typ}} = 15.7 \text{ mW}$ \\
 & $P_{\text{max}} = 76.6 \text{ mW}$ \\ \hline
RMS Jitter & $J_{\text{typ}} = 550 \text{ fs}$ \\
 & $J_{\text{max}} = 814 \text{ fs}$ \\ \hline
Duty Cycle of Sampling Clock & $46.3 - 54.8\%$ \\ \hline
Charge Pump Current & $I_{cp} = 200 \ \mu A$ \\ \hline
Characteristic Damping Time & \multirow{2}{*}{$\tau = 1 - 2 \ \mu s$} \\
Scale & \\ \hline
Phase Margin & $50.13^\circ$ \\ \hline
\end{tabular}
\caption{PSEC6 10.24 GHz PLL simulation parameters.}
\label{tab: PLL specifications}
\end{table}

\section{Conclusions}

\begin{figure*}[t]
    \centering
    \includegraphics[width=0.4\linewidth, angle=90]{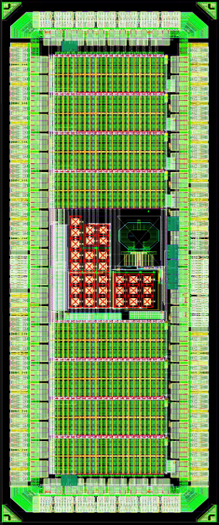}
    \caption{\textbf{PSEC6 Layout}: Layout of PSEC6 with guard ring, I/O pads, ESD devices, channels, digital block, and PLL shown.
    There are eight channels containing SCAs, four on the left half and four on the right half.
    The PLL is placed in the middle of the chip to minimize the travel distance of the sampling clock.
    The PLL contains the inductor in the upper left corner and the LPF Metal-Insulator-Metal (MIM) capacitors in the bottom right corner.
    The power domains of the digital and analog I/O are separated along the I/O ring corners.}
    \label{fig: PSEC6 layout}
\end{figure*}

PSEC6 has been taped out as of April 2026, and is scheduled for delivery back from TSMC in July 2026.
Simulations predict a 4.0 GHz analog input bandwidth and 20 mW per channel; the 10.24 GHz PLL achieves 550 fs RMS jitter (typical corner) at 15.7 mW.
The final layout of PSEC6 is shown in Figure \ref{fig: PSEC6 layout}.
The physical testing of PSEC6 will happen when the chip arrives.
Designs of the test setup and applications to experiment for PSEC6 are currently under development.

\clearpage
\section*{Acknowledgments}

We would like to thank Eric Oberla and Evan Angelico for their foundational work on the PSEC project.
We would also like to thank Maresa Wynd and Fukun Tang of the Electronics Design Group at the University of Chicago for their technical help. 
Finally, we would like to acknowledge Stefan Ritt for his insights into the applications and challenges of fast timing. 




  \bibliographystyle{elsarticle-num-names} 
  \bibliography{main}

@INPROCEEDINGS{diffFF,
  author={Jiren Yuan and Svensson, C.},
  booktitle={1996 Symposium on VLSI Circuits. Digest of Technical Papers}, 
  title={New {TSPC} latches and flip-flops minimizing delay and power}, 
  year={1996},
  volume={},
  number={},
  pages={160-161},
  doi={10.1109/VLSIC.1996.507754}}

@inproceedings{heydari2003cml,
  author    = {Heydari, Payam and Mohanavelu, Ravi},
  title     = {Design of Ultra High-Speed {CMOS} {CML} Buffers and Latches},
  booktitle = {2003 IEEE International Symposium on Circuits and Systems ({ISCAS})},
  year      = {2003},
  volume    = {2},
  pages     = {II--208--II--211},
  isbn      = {0-7803-7761-3},
}

@inproceedings{rhee1999chargepump,
  author    = {Rhee, Woogeun},
  title     = {Design of High-Performance {CMOS} Charge Pumps in Phase-Locked Loops},
  booktitle = {1999 IEEE International Symposium on Circuits and Systems ({ISCAS})},
  year      = {1999},
  volume    = {2},
  pages     = {II--545--II--548},
  isbn      = {0-7803-5471-0},
}

@inproceedings{mohanavelu2004divider,
  author    = {Mohanavelu, Ravindran and Heydari, Payam},
  title     = {A Novel Ultra High-Speed Flip-Flop-Based Frequency Divider},
  booktitle = {2004 IEEE International Symposium on Circuits and Systems ({ISCAS})},
  year      = {2004},
  volume    = {4},
  pages     = {169--172},
  doi       = {10.1109/ISCAS.2004.1328967},
}

@article{razavi2016tspc,
  author    = {Razavi, Behzad},
  title     = {{TSPC} Logic [{A Circuit for All Seasons}]},
  journal   = {{IEEE} Solid-State Circuits Magazine},
  year      = {2016},
  volume    = {8},
  number    = {4},
  pages     = {10--13},
  month     = {Fall},
  doi       = {10.1109/MSSC.2016.2603228},
}

@article{mansuri2002pfd,
  author    = {Mansuri, Mozhgan and Liu, Dean and Yang, Chih-Kong Ken},
  title     = {Fast Frequency Acquisition Phase-Frequency Detectors for {Gsamples/s}
               Phase-Locked Loops},
  journal   = {{IEEE} Journal of Solid-State Circuits},
  year      = {2002},
  volume    = {37},
  number    = {10},
  pages     = {1331--1334},
  month     = {October},
}

@inproceedings{sharma2021pfdreview,
  author    = {Sharma, Jyoti and Varma, Tarun and Boolchandani, Dharmendar},
  title     = {A Brief Review of the Various Phase-Frequency Detector Architectures},
  booktitle = {2021 IEEE International Symposium on Smart Electronic Systems ({iSES})},
  year      = {2021},
  pages     = {74--78},
  doi       = {10.1109/iSES52644.2021.00028},
  isbn      = {978-1-7281-8753-2},
}

@inproceedings{braga2024cryopll,
  author    = {Braga, Davide and Das, Kushal and Dick, Neil and Dyer, Ken
               and England, Troy and Fahim, Farah and Holm, Scott and Lu, Ping
               and Moini, Alireza and Nolet, Frederic and Rubinov, Paul
               and Sanderson, Len and Thompson, Barry and Wang, Xiaoran},
  title     = {Design of a Low-Jitter 10~{GHz} {PLL} for a 12-bit 10-{GSPS}
               Cryogenic {ADC} for Quantum Readout in 22{FDX}},
  booktitle = {Coordinating Panel for Advanced Detectors Workshop ({CPAD} 2024)},
  year      = {2024},
  month     = {November},
  address   = {University of Tennessee, Knoxville, TN},
  note      = {Fermilab Report FERMILAB-SLIDES-24-0005-ETD-PPD},
}

@inproceedings{wang2024cryopll,
  author    = {Wang, Xiaoran and Braga, Davide and England, Troy and Das, Kushal
               and Dick, Neil and Fahim, Farah and Holm, Scott and Moini, Alireza
               and Nolet, Frederic and Rubinov, Paul and Sanderson, Len
               and Saxena, Yash},
  title     = {A 10~{GHz} Low-Jitter Cryogenic Phase-Locked Loop ({PLL})
               for Quantum Applications},
  booktitle = {Coordinating Panel for Advanced Detectors Workshop ({CPAD} 2024)},
  year      = {2024},
  month     = {November},
  day       = {19},
  address   = {University of Tennessee, Knoxville, TN},
  note      = {Fermilab Report FERMILAB-SLIDES-24-0310-ETD},
}

@misc{ritt2011analog,
  author       = {Ritt, Stefan},
  title        = {The Role of Analog Bandwidth and Signal-to-Noise in Timing
                  for Waveform Digitizing},
  howpublished = {Talk presented at the Timing Workshop, Chicago},
  year         = {2011},
  month        = {April},
  url = {https://psec.uchicago.edu/workshops/fast_timing_conf_2011/system/docs/18/original/ritt_analog_bw.ppt
},
}

@article{park2025psec5,
  author    = {Park, Jinseo and Angelico, Evan and Arzac, Andrew and Braga, Davide
               and Datta, Ahan and England, Troy and Ertley, Camden and Fahim, Farah
               and Frisch, Henry J. and Heintz, Mary and Oberla, Eric
               and Pastika, Nathaniel J. and Rico-Aniles, Hector D. and Rubinov, Paul M.
               and Wang, Xiaoran and Yeung, Yui Man Richmond and Zimmerman, Tom N.},
  title     = {Design of an 8-channel 40~{GS/s} 20~{mW/Ch} Waveform Sampling {ASIC}
               in 65~nm {CMOS}},
  journal   = {Nuclear Instruments and Methods in Physics Research Section {A}:
               Accelerators, Spectrometers, Detectors and Associated Equipment},
  year      = {2025},
  volume    = {1072},
  pages     = {170165},
  doi       = {10.1016/j.nima.2024.170165},
  eprint    = {2407.09575},
  archivePrefix = {arXiv},
}

@inproceedings{datta2024cpad,
  author    = {Datta, Ahan and Angelico, Evan and Arzac, Andrew and Braga, Davide
               and England, Troy and Frisch, Henry J. and Heintz, Mary and Oberla, Eric
               and Pastika, Nathaniel J. and Park, Jinseo and Rico-Aniles, Hector D.
               and Rubinov, Paul M. and Wang, Xiaoran and Yeung, Y. M. Richmond},
  title     = {Design and Implementation for the Digital Block of {PSEC5}},
  booktitle = {Coordinating Panel for Advanced Detectors Workshop ({CPAD} 2024),
               Session RDC4: Readout and {ASICs}},
  year      = {2024},
  month     = {November},
  day       = {20},
  address   = {University of Tennessee, Knoxville, TN},
  url       = {https://indico.phy.ornl.gov/event/510/contributions/2364/},
}

@inproceedings{datta2026aps,
  author    = {Datta, Ahan and Angelico, Evan and Arzac, Andrew and Chen, Gordon
               and England, Troy and Frisch, Henry J. and Gehl, Nathan and Heintz, Mary
               and Kim, Sumin and Lalich, Ava and Pastika, Nathaniel J. and Park, Jinseo
               and Rico-Aniles, Hector D. and Rubinov, Paul M. and Yeung, Y. M. Richmond},
  title     = {10.24~{GHz} {PLL} for {PSEC6}, a 40~{GSa/s} Waveform Sampling {ASIC}
               in 65~nm {CMOS}},
  booktitle = {{APS} Global Physics Summit 2026, Session {APR-V88}},
  year      = {2026},
  month     = {March},
  day       = {19},
  url       = {https://summit.aps.org/smt/2026/events/APR-V88/5},
}

\end{document}